\documentclass[english,epj]{svjour}
\usepackage{amssymb}
\usepackage{amsmath}
\usepackage{mathtools}
\usepackage{eucal}
\usepackage{bm}
\usepackage{babel}
\usepackage{slantsc}
\usepackage{array}
\usepackage{hyperref}
\usepackage[numbers, sort&compress]{natbib}
\usepackage{url}
\def\be{\begin{equation}}
\def\ee{\end{equation}}
\def\ba{\begin{eqnarray}}
\def\ea{\end{eqnarray}}
\def\bsu{\begin{subequations}}
\def\esu{\end{subequations}}

\def\a{\alpha}
\def\b{\beta}
\def\g{\gamma}     
\def\d{\delta}

\def\m{\mu}
\def\n{\nu}
\def\o{\omega}   \def\O{\Omega}

\def\t{\tau}

\def\lab{\label}
\def\pd{\partial}
\def\le{\left}
\def\ri{\right}
\def\mm{{\mathtt f}}

\begin{document}
\title{Local gravitational physics of the Hubble expansion}
\subtitle{Einstein's equivalence principle in cosmology}
\author{Sergei M. Kopeikin\inst{1,2,}\thanks{E-mail: kopeikins@missouri.edu}}
\institute{$^1$Department of Physics \& Astronomy, University of Missouri, 322 Physics Bldg., Columbia, MO 65211, USA\\$^2$
Siberian State Geodetic Academy, 10 Plakhotny Street, Novosibirsk 630108, Russia}
\date{Received: 12-12-2014   /  Accepted: 10-01-2015 }
\abstract{
We study physical consequences of the Hubble expansion of Friedmann-Lema\^itre-Robertson-Walker (FLRW) manifold on measurement of space, time and light propagation in the local inertial frame. We use results of this study to analyse the solar system radar ranging and Doppler tracking experiments and time synchronization.
FLRW manifold is covered by the coordinates $(t,y^i)$, where $t$ is the cosmic time coinciding with the proper time of the Hubble observers and identified with the barycentric coordinate time (TCB) used in ephemeris astronomy.
We introduce local inertial coordinates $x^\a=(x^0,x^i)$ in the vicinity of a world line of a Hubble observer with the help of a special conformal transformation that respects the local equivalence between the tangent and FLRW manifold. The local inertial metric is Minkowski flat and is materialized by the congruence of time-like geodesics of static observers being at rest with respect to the local spatial coordinates $x^i$. The static observers are equipped with the ideal clocks measuring their own proper time  which is synchronized with the cosmic time $t$ measured by the Hubble observer. 
We consider geodesic motion of test particles and notice that the local coordinate time $x^0=x^0(t)$ taken as a parameter along the world line of particle, is a function of the Hubble's observer time $t$.
This function changes smoothly from $x^0=t$ for a particle at rest (observer's clock), to $x^0=t+(1/2)Ht^2$ for photons, where $H$ is the Hubble constant. Thus, motion of a test particle is non-uniform when its world line is parametrized by the cosmic time $t$. NASA JPL Orbit Determination Program operates under assumption that spacetime is asymptotically flat which presumes that motion of light (after the Shapiro delay is excluded) is uniform with respect to the time $t$ but it does not comply with the non-uniform motion of light on cosmological manifold. For this reason, the motion of light in the solar system analysed with the Orbit Determination Program appears as having a systematic {\it blue} shift of frequency, of radio waves circulating in the Earth-spacecraft radio link.
The magnitude of the anomalous blue shift of frequency is proportional to the Hubble constant $H$ that may open an access to the measurement of this fundamental cosmological parameter in the solar system radiowave experiments.
}
\maketitle

%\tableofcontents

\section{Introduction}

Modern physics is intensively looking for the unified field theory that might explain the origin of the universe and the underlying fundamental nature of spacetime and elementary particles  \cite{baryshev_2012ASSL}. This work requires deeper understanding of the theoretical and experimental principles of general relativity. It is challenging to find a new type of experiments that broaden the current knowledge. An appealing problem is to examine a presumable link between the local gravitational phenomena and the global cosmological expansion of the universe that is to test the foundational basis of the Einstein equivalence principle (EEP) in application to a conformal cosmological metric with a time-dependent scale factor. 

EEP in general relativity is universally valid because gravitational field in general relativity has a pure geometric nature. It is always mathematically possible to find a local diffeomorphism which reduces any global metric to a Minkowski metric in a sufficiently small neighborhood of a time-like world line of observer if tidal forces are neglected. This mathematical fact was a clue that led Einstein to formulation of his general principle of relativity (also known as the principle of covariance \cite{kopeikin_2011book}) and, later on, to the discovery of general relativity as a physical theory of gravitational field \cite{einstein1961relativity,Einstein_P1916}. 

The apparent mathematical nature of EEP caused some physicists to deny its physical significance \cite{Norton_1993}. The present paper neither shares this extremal point of view nor confronts the solid mathematical foundation of EEP. We focus on physical aspects of EEP, namely, 
\begin{enumerate}
\item comparison of the inertial motion of test particles on cosmological manifold considered from the local point of view of a Hubble observer,
\item derivation of experimental consequences that can be used for testing the Hubble law in the local solar system experiments.  
\end{enumerate} 

So far, all gravitational experiments in the solar system have been interpreted under a rather natural assumption that the background spacetime geometry is asymptotically-flat \cite{will_1993} covered by coordinates $(t,y^i)$ with the background Minkowski metric
\be\lab{n7dl}
ds^2=-dt^2+\d_{ij}dy^i dy^j\;,
\ee
where Latin indices $i,j,k,..$ take values $1,2,3$, $\d_{ij}={\rm diag}(1,1,1)$ - the unit matrix, and we have used a convention for the speed of light, $c=1$. The time $t$ entering the metric \eqref{n7dl} is identified with the barycentric coordinate time (TCB) of the solar system according to the IAU 2000 resolutions \cite{2003AJ....126.2687S}.

On the other hand, theoretical and observational cosmology postulates that the background spacetime is described by the Friedmann-Lema\^itre-Robertson-Walker (FLRW) metric 
\be\lab{1}
ds^2=-dt^2+{R}^2(t)\le(1+\frac14 kr^2\ri)^{-2}\d_{ij}dy^idy^j\;,
\ee
where $t$ is the universal cosmic time, $y^i$ are the global isotropic coordinates,  $k=\{-1,0,+1\}$ defines a curvature of space, and the scale factor $R(t)$ is a function of time found by solving Einstein's equations \cite{mukh_book,weinberg_2008}. The cosmic time $t$ is the proper time measured by observers having fixed spatial coordinates $y^i$. Therefore, it is exactly the same as time $t$ in the flat metric \eqref{n7dl} and identifies with TCB of fundamental astronomy in the solar system \cite{2003AJ....126.2687S,kopeikin_2011book,2013strs.book.....S}.

In what follows, we admit $k=0$ in accordance with observations \cite{wmap_2011} and limit ourselves with the linearized Hubble approximation. In other words, we consider only terms being linear with respect to the Hubble constant $H$ and neglect all terms that are quadratic with respect to $H$ or proportional to its time derivative $\dot H\sim H^2$. We shall also neglect post-Newtonian gravitational effects of the solar system which must be included to the realistic data analysis of observations in the solar system. These effects are described in \cite{kopeikin_2011book,2013strs.book.....S}, and can be easily accounted for by superposition, if necessary. 

FLRW metric (\ref{1}) is not asymptotically-flat and has a non-vanishing spacetime curvature tensor $ R_{\a\b\g\d}$ where the Greek indices $\a,\b,\g,...$ take values $0,1,2,3$. Nonetheless, the Weyl tensor of FLRW metric, $ C_{\a\b\g\d}\equiv 0$. Hence, (\ref{1}) can be reduced to a conformally-flat metric for any value of space curvature $k$ \cite{2007JMP....48l2501I}. When $k=0$, it is achieved by transforming the cosmological time $t$ to a conformal time, $\eta=\eta(t)$, defined by an ordinary differential equation
\be\lab{oka1}
dt=a(\eta)d\eta\;,
\ee
where the scale factor, $a(\eta)\equiv R[t(\eta)]$. The time transformation (\ref{oka1}) brings the cosmological metric (\ref{1}) into the conformally-Minkowskian form
\be\lab{q1}
ds^2=a^2(\eta)\mm_{\a\b}dy^\a dy^\b\;,
\ee
where $y^\a=(y^0,y^i)=\le(\eta,y^i\ri)$ are the global conformal coordinates, $\mm_{\a\b}={\rm diag}(-1,1,1,1)$ is the Minkowski metric. 

According to Einstein's general relativity and the definition of FLRW metric,
the cosmological time $t$ is a physical proper time of the Hubble observer and can be measured with the help of the observer's atomic clock while the conformal time $\eta$ is a convenient coordinate parameter which is calculated from the clock's reading but cannot be measured directly \cite{mukh_book,weinberg_2008}. Typically, the cosmological metric (\ref{q1}) is applied to describe the properties of spacetime on the scale of galaxy clusters and larger. On small scales of the size of the Milky Way, the solar system and terrestrial lab, the background spacetime is believed to be flat with any cosmological effect being strongly suppressed. Nevertheless, the question remains open: if we admit FLRW metric to be valid on any scale, can the cosmological expansion be detected in local gravitational experiments? 

Following \cite{2010RvMP...82..169C,Kopeikin_2012eph} we postulate that FLRW metric (\ref{q1}) is a physical metric not only in cosmology but for the  description of the local physics as well. It describes the background spacetime geometry in the global coordinates $y^\a$ on all scales spreading up from the cosmological horizon to the solar system and down to a local observer. The small parameter in the approximation scheme used in the present paper, is the product of the Hubble constant, $H$, with the interval of time used for physical measurements. All non-linear terms of the quadratic order with respect to the small parameter (formally, the terms being quadratic with respect to $H$) will be systematically neglected because of their smallness. 

We introduce the reader to the concepts associated with the Einstein equivalence principle in section \ref{eepr} and discuss construction of the local inertial coordinates in section \ref {preq}. The inertial frame is built in sections \ref{sec4}. Light geodesics in local coordinates are derived in section \ref{opm123}. We solve these equations in section \ref{rr1} and employ them for investigation of observability of cosmological effects in the solar system. Discussion is provided in section \ref{opwd}.

\section{Einstein's principle of equivalence}\lab{eepr}

A thorough  treatment of the local astronomical measurements on cosmological manifold inquires a scrutiny re-examination of Einstein's equivalence principle (EEP) which states: ``In a given gravitational field, the outcome of {\it any} local, non-gravitational experiment is independent of the freely-falling experimental apparatus' velocity, of where and when in the gravitational field the experiment is performed and of experimental technique applied'' \cite{Rohrlich1963169}. Mathematical interpretation of EEP suggests universality of local geometry in the sense that at each point on a spacetime manifold with an arbitrary gravitational field, it is possible to chose the local inertial coordinates such that, within a sufficiently small region of the point in question, {\it all} laws of nature take the same form as in non-accelerated Cartesian coordinates \cite{Harvey1964383,kopeikin_2011book}. 

EEP is applicable in general relativity to any kind of spacetime manifold, in particular, to the manifold of FLRW universe \cite{2010RvMP...82..169C} which is described by metric (\ref{q1}). We noticed \cite{Kopeikin_2012eph,kopeikin_2013} that due to the expanding nature of space in FLRW manifold, the parametric description of the propagation of light given in local inertial coordinates in terms of the proper time of observer, differs from that in the Cartesian coordinates of flat spacetime. Let us consider a Hubble observer who is at the origin of a local inertial coordinates (LIC), $x^\a=(x^0,x^i)$. Physical metric $ g_{\a\b}\equiv a^2(\eta)\mm_{\a\b}$, given by (\ref{q1}) in the global coordinates, is reduced to the Minkowski metric, $\mm_{\a\b}$, at the origin of LIC
with the affine connection being nil, $\Gamma^\a_{\m\n}(x)=0$, on the observer's world line \cite{hongya:1920,hongya:1924,2005ESASP.576..305K,2007CQGra..24.5031M}. EEP asserts that the worldlines of freely falling (electrically-neutral) test particles and photons are geodesics of the physical metric $ g_{\a\b}$ with an affine parametrization. Because the affine connection is nil in LIC, it presumes that the geodesic equations of motion of {\it all} test particles -- massive and massless -- can be written down as follows 
\be\lab{bq1}
\frac{d^2x^\a}{d\sigma^2}=0\;,
\ee
where $\sigma$ is the affine parameter along the geodesic. Equation (\ref{bq1}) neglects the tidal (caused by the Riemann curvature of FLRW spacetime) effects \cite{nizim} which produce terms of the order of $H^2$ which we discard. Solving \eqref{bq1} for time component shows that the parameter $\sigma$ can be chosen equal to the coordinate time $x^0$ of LIC, that is $\sigma=x^0$.

Local coordinate time, $x^0$, must be further operationally connected to the proper time $t$ measured in LIC by the central Hubble observer. The time $x^0$ is often identified with the proper time of observer $t$ but one must keep in mind that this identification is true only for static observers being at rest with respect to the central Hubble observer. In general, the local coordinate time $x^0$ is a non-linear function of $t$ on worldlines of moving test particles. Therefore, changing the affine parameter $\sigma$ to the non-affine (but directly measurable) parameter $t$ brings \eqref{bq1} to the following form
\be\lab{bq1q2}
\frac{d^2x^\a}{dt^2}=\frac{dx^\a}{dx^0}\frac{d^2x^0}{dt^2}\;.
\ee 

The cosmic time $t$ coincides with the proper time of the central Hubble observer in the absence of any gravitational perturbations caused by massive bodies of the solar system like Sun and planets. Real experiments demand to include the effect of gravitational field of the solar system on time transformations but they are well-known and can be easily taken into account \cite{kopeikin_2011book}. In ephemeris astronomy the time $t$ is identified with the barycentric coordinate time (TCB) which is considered as a uniform global time scale. Equations for transformation of the proper time $\t$ of any observer within the solar system to TCB are given by the corresponding IAU resolutions \cite{2003AJ....126.2687S}. This transformation shows that $\t$ differs from $t$ by small relativistic terms which are not essential for further discussion (see \cite{kopeikin_2011book} for more detail). 

NASA JPL Orbit Determinantion Program that is used for spacecraft navigation and calculating planetary and lunar ephemerides \cite{odprogram} assumes that for any particle including photons, $x^0=t$. It means that the right side of \eqref{bq1q2} is postulated to be nil
\be\lab{bq1a}
\frac{d^2x^\a}{dt^2}=0\;.
\ee
This equation yields the photon's world line $x^0=t,\;x^i=x^i_0+k^it$, where $k^i$ is a unit vector in the direction of the photon's propagation and we assumed that light passes through the point $x^i_0$ of LIC at instant $t=0$ which fixes the integration constants \footnote{Realistic measurement requires accounting for the post-Newtonian relativistic corrections in (\ref{bq1a}) which is beyond the scope of the present paper but can be found, for example, in \cite{kopeikin_2011book,2013strs.book.....S}.}. It establishes a linear relationship between the spatial coordinates $x^i$ of the photon and the proper time $t$ of the observer at the origin of LIC, which is a directly measurable quantity (after accounting for the IAU time transformations). 

Equation (\ref{bq1a}) does not show the presence of the Hubble constant, $H$. It led scientists to believe that EEP cancels out all cosmological effects of the linear order of $O(H)$ that prevents astronomers to observe them in the solar system' experiments \cite{2010RvMP...82..169C}. However, equation \eqref{bq1a} and its solution are incomplete as the relation between $x^0$ and $t$ is not a linear function of time so that light does not propagate uniformly with constant velocity. This non-uniform propagation of light in the local coordinates may look as anomaly and violation of EEP for photons but this is just a mathematical consequence of the geometric expansion of space in FLRW universe. This effect makes possible measurement of the Hubble expansion in the solar system in the local experiments like the Doppler tracking of spacecraft in deep space.  

\section{The local inertial coordinates}\lab{preq}

In order to interpret the local astronomical measurements (like radar ranging, spacecraft Doppler tracking, etc.) we have to build the local inertial coordinates (LIC) in the neighbourhood of a time-like world line of observer. We focus on building LIC in the vicinity of a Hubble observer which is by definition has constant spatial coordinates $y^i={\rm const.}$ of the FLRW metric and moves along a time-like geodesic worldline \cite{waldorf}. Real observers move with respect to the Hubble flow and experience gravitational forces from the massive bodies of the solar system. Therefore, construction of LIC for a real observer requires to work out additional coordinate transformations which are known and can be found in \cite{kopeikin_2011book,2013strs.book.....S} so that they are not a matter of concern of the present paper.  

Let us put the Hubble observer at the origin of LIC, $x^i=0$, which worldline coincides, then, with the time-like geodesic of the observer. The Hubble observer carries out an ideal clock that measures the parameter of the observer's worldline which is the observer's proper time $t$. The proper time $t$ of the Hubble observer coincides with the cosmological coordinate time $t$ in (\ref{1}). EEP suggests that in a small neighbourhood of the worldline of the observer (called a tangent spacetime) there exists a {\it local} diffeomorphism from the global, $y^\a$, to local, $x^\a$, coordinates such that the physical metric $ g_{\a\b}(y)=a^2(\eta)\mm_{\a\b}$ is transformed to the Minkowski metric, $\mm_{\a\b}$, as follows
\be\lab{c2}
a^2(\eta)\mm_{\m\n}\frac{\pd y^\m}{\pd x^\a} \frac{\pd y^\n}{\pd x^\b}=\mm_{\a\b}\;,
\ee
where all tidal terms of the order of $O(H^2)$ have been omitted as negligibly small. In the tangent spacetime where (\ref{c2}) is valid, the physical spacetime interval (\ref{q1}) written down in LIC, reads
\be\lab{act6}
ds^2=\mm_{\a\b}dx^\a dx^\b\;,
\ee
where $\mm_{\a\b}$ is understood as the physical metric $ g_{\a\b}(x)$ expressed in the local coordinates.    

Equation (\ref{c2}) looks similar to the special conformal transformation establishing a conformal isometry of the Minkowsky metric  \cite{1966PhRv..150.1183K,cft_intro}
\be\lab{c5}
\O^{2}(x)\mm_{\m\n}\frac{\pd y^\m}{\pd x^\a} \frac{\pd y^\n}{\pd x^\b}=\mm_{\a\b}\;,
\ee
where
\ba\lab{c6}
\O(x)&=&\mm_{\a\b}\le(b^2x^\a-b^\a\ri)\le(b^2x^\b-b^\b\ri)b^{-2}\\\nonumber
&=&1-2b_\a x^\a+b^2 x^2\;,
\ea
is a conformal factor, $b^\a$ is a constant four-vector yet to be specified, $x^2\equiv\mm_{\a\b}x^\a x^\b$, and
$b^2\equiv\mm_{\a\b}b^\a b^\b$. The special conformal transformation
includes inversions and translations, and is defined by equation \cite{1962AnP...464..388K,cft_intro}
\be\lab{mu7a}
\frac{y^\a}{y^2}=\frac{x^\a}{x^2}-b^\a\;,
\ee
that is equivalent to
\be\lab{c4}
y^\a=\frac{x^\a-b^\a x^2}{\O(x)}\;.
\ee
All operations of rising and lowering indices in the above equations are completed with the Minkowski metric $\mm_{\a\b}$.  

Let us assume for simplicity that the origin of LIC, $x^i=0$, coincides with the point having the global spatial coordinates, $y^i=0$. As the background manifold is assumed to be analytic, equation
(\ref{c2}) should match (\ref{c5}) in a small neighbourhood of the origin of the LIC. The matching can be achieved by demanding the scale factor of the FLRW metric, $a(\eta(x))=\Omega(x)$. This equality is 
valid in arbitrary cosmological model if we discard the curvature terms being proportional to $\sim H^2$ and/or $\dot H$. Indeed, for small values of the conformal time $\eta$ we have,
\be\lab{mn5va}
a(\eta)=a(0)+a'(0)\eta+\frac12 a''(0)\eta^2+O(\eta^3)\;,
\ee
where we assume that the present epoch corresponds to $\eta=0$ in the conformal time, and the prime
denotes the time derivative, $a'=da/d\eta$, etc. We normalize the scale factor at the present epoch to $a(0)=1$. Then, at the present epoch the Hubble
constant $H=a'(0)$. The second time derivative of the scale factor, $a''=H'+2H^2$, and we drop it off as being negligibly small. Assuming that the constant vector, $b^\a=O(H)$, we approximate the conformal factor, $\O(x)=1-2b_\a x^\a$ by neglecting terms of the order of $b^2$. Taking into account that (\ref{c4}) yields $\eta=x^0+O(b)$, and equating $\O(x)$ in (\ref{c6}) to the Taylor expansion \eqref{mn5va} of the scale
factor $a(\eta)$, we find out that matching yields vector $b^\a=(H/2)u^\a=(H/2,0,0,0)$. It is directed along the four-velocity $u^\a=(1,0,0,0)$ of the Hubble observer, and is time-like. 

The reader may notice that the special conformal transformation has a singular point, $x^\a=-b^\a/b^2$, that goes over to $t=2/H$. It means that the special conformal diffeomorphism (\ref{c4}) is approximately limited in time domain by the Hubble time, $T_H=1/H$, calculated for the present value of the Hubble parameter $H\simeq 2.3\times 10^{-18}$ s$^{-1}$. However, because LIC have been derived under assumption that the series \eqref{mn5va} is convergent, $Ht\ll 1$, the period of time for which the local inertial frame is really valid is much smaller than the Hubble time and is given by, $t\ll T_H$. Because of this limitation imposed on the time of applicability of the local frame, the local coordinates are also bounded in space by the radius, $r\ll R_H$, where $R_H=cT_H$ is the Hubble radius of the universe. The conclusion of this paragraph is that the LIC can be employed only for sufficiently close objects in the universe with the redshift factor $z\ll 1$ which excludes the most distant quasars and galaxies. Therefore, the formalism of the present paper is not applicable to the discussion of global cosmological properties and/or effects like the red shift of quasars. More stringent results on the domain of applicability of the local inertial coordinates in cosmology can be found in \cite{klein_2011AnHP}.

The matching ensures that LIC can be constructed in the linearised Hubble approximation from the global coordinates, $y^\a$, by means of the special conformal transformation
(\ref{c4}) that respects EEP as the matching procedure demonstrates. In what follows, we accept the equalities, $\O(x(\eta))=a(\eta)$ that are valid in the linearised Hubble approximation. Moreover, we work in the vicinity to the present epoch where $a(t)=1+Ht+O(H^2t^2)$, and in this approximation we are allowed to use $t=\eta$ in terms which are proportional to $H$. It means we can equate $a(\eta)=a(t)$.  This brings the transformation \eqref{c4} to the following form 
\ba\lab{c4dd}
x^0&=&a(\eta)\left[\eta-\frac{H}2\left(\eta^2-{\bm y}^2\right)\right]\;,\\\lab{b6ft}
x^i&=&a(\eta)y^i\;,
\ea
where ${\bm y}=y^i$, ${\bm y}^2=\d_{ij}y^iy^j$, and all residual terms of the order of $H^2$ have been discarded. Expansion of the scale factor $a(\eta)$ and conformal time $\eta$ in terms of the cosmic time $t$ yields yet another form of the transformation from global to local coordinates in cosmology
\ba\lab{c4rr}
x^0&=&t+\frac{H}2{\bm y}^2\;,\\\lab{c4sf}
x^i&=&(1+Ht)y^i\;,
\ea
where we have again neglected all residual terms of the quadratic order in $H$.

The special conformal transformation (\ref{c4}) and its approximate expressions \eqref{c4dd}--\eqref{c4sf} extends the list of transformations to LIC in cosmology found by other researchers \cite{hongya:1924,2005ESASP.576..305K,2007CQGra..24.5031M,klein_2011AnHP}.

\section{The local inertial frame}\lab{sec4}

The local inertial coordinates, $x^\a$, are mathematical functions on FLRW manifold which have no immediate physical meaning unlike the Cartesian coordinates in Euclidean space. To make the local coordinates physically meaningful they should be further specified and operationally connected with measuring devices (clocks, rulers) of a set of some reference observers. This materialization yields access to the local inertial frame. The corresponding relations between the measuring tools and the local coordinates are known in differential geometry as inertial (or projective) structure \cite{eps_1972}. The Minkowski form of the physical local metric (\ref{act6}) suggests that LIC can be associated with the Gaussian normal coordinates based on the congruence of time-like geodesics of (electrically-neutral) test particles being at rest with respect to LIC \cite{waldorf}. 

The first step, is to find relation between the coordinate time $x^0$ and the proper time $t$ of the Hubble observer at the origin of LIC. Because the spacetime interval, $ds^2=-d(x^0)^2$ for $x^i=0$, and $ds^2=-dt^2$ by the definition of the proper time \cite{waldorf}, we come to the conclusion that $x^0=t$ on the worldline of the origin of LIC. The grid of the Gaussian coordinates start from the initial hypersurface, $t=0$, that is orthogonal to the worldline of the Hubble observer. We identify the spatial Gaussian coordinates with the orthogonal (in the Euclidean sense) spatial coordinates $x^i$ of LIC on the initial hypersurface. Extension of the spatial coordinates from the initial hypersurface to arbitrary value of the time coordinate $x^0\equiv t$ is performed by means of time-like geodesics. The Christoffel symbols of the local metric (\ref{act6}) are nil in a neighbourhood of the origin of LIC in accordance with diffeomorphism (\ref{c4}) by which LIC were introduced.  Because all Christoffel symbols are nil, the time-like worldlines of particles having constant spatial coordinates, $x^i={\rm const.}$, are geodesics given by (\ref{bq1}). The proper time of the particle with the constant spatial coordinate $x^i$ coinsides with the time coordinate $x^0$ which was identified with the proper time of the Hubble observer. Hence, the parameter $\sigma$ in (\ref{bq1}) can be identified with the proper time $t$ as well. After that equation (\ref{bq1}) describing worldlines of the static observers takes on the following simple form,  
\be\lab{bqq2}
\frac{d^2x^\a}{dt^2}=0\;.
\ee 

The meaning of time-like geodesic equation (\ref{bqq2}) is as follows.
The world lines $x^\a=\le\{x^0=t,x^i={\rm const.}\ri\}$ are identified with the network of static reference observers which play a fundamental role in local physical measurements. We admit that each static observer is equipped with an ideal (atomic) clock measuring their proper time which coincides with a time-like parameter, $x^0$, along the observer's worldline. Solving (\ref{bqq2}) reveals that $x^0\equiv t$ is the proper time of the Hubble observer located at the origin of LIC. We assume that the ideal clocks of the static observers are synchronized. It can be done with Einstein's procedure of exchanging light signals as we will confirm in section \ref{bub3}. 

The Gaussian normal coordinates form a local inertial frame that is used for doing local physical measurements of time and space along time-like world lines of static observers and on space-like hypersurfaces of constant time. The frame is defined operationally in terms of the proper time of the ideal clocks and rigid rulers. The rulers are made of an ordinary matter which rigidity is defined primarily by the chemical bonds having an electromagnetic origin. We have proved \cite{Kopeikin_2012eph} that in the linearized Hubble approximation the electromagnetic (Coulomb) forces in an expanding universe remain the same as in a flat spacetime. For this reason, the rigid rulers and rods are not subject to the cosmological expansion and can serve for physical materialization of LIC. Another physical realization of the local Gaussian coordinates is achieved by the celestial ephemerides of the solar system bodies since their orbits are not affected by the Hubble expansion either \cite{2010RvMP...82..169C,Kopeikin_2012eph}.   

\section{The light geodesics}\lab{opm123} 

The most precise measurements of spacetime events are made with electromagnetic waves and light \cite{2006LRR.....9....3W}. Therefore, we have to solve equations of light geodesics \eqref{bq1q2} parametrized with the proper time $t$ of the central observer which is directly measurable quantity.  First of all, we need to evaluate the right side of \eqref{bq1q2}. Function $x^0$ taken on the light cone, where $\eta^2-{\bm y}^2=0$, is given in terms of the conformal time $\eta$ by \eqref{c4dd} or, more exactly, $x^0=a(\eta)\eta$. Since the conformal time $\eta$ and the cosmic time $t$ are related by 
\be\lab{h6f}
t=\int a(\eta)d\eta=\eta+\frac{H}2\eta^2+O(H^2)\;,
\ee 
and the cosmic time coincides with the proper time $t$ of the central Hubble observer, we get on the light geodesic
\be\lab{yu6v}
x^0=t+\frac{H}2t^2\;,
\ee 
which can be also obtained directly from \eqref{c4rr} after making use of equation of light geodesics for ${\bm y}^2\simeq t^2$ in the term being proportional to $H$.
Taking the second derivative from $x^0$ in \eqref{yu6v} yields $d^2x^0/dt^2=H$. Hence, equation of light geodesics \eqref{bq1q2} takes on in the local coordinates the following form
\be\lab{v3x7}
\frac{d^2x^\a}{dt^2}=H\frac{dx^\a}{dt}\;,
\ee
where we have made use of a legitimate approximation, $dx^\a/dx^0=dx^\a/dt$, in the right side of \eqref{v3x7}.

Equation (\ref{v3x7}) predicts the existence of a cosmological force in the tangent space of FLRW universe, exerted on a freely-falling photon. It should not be misinterpreted as a violation of general relativity or Newtonian gravity like the ``fifth force'' \cite{1992Natur.356..207F} or whatever else. Equation (\ref{v3x7}) is a direct consequence of general relativity applied along with the cosmological principle stating that the global cosmological time $t$ is identical with the proper time measured by the Hubble observer. It explains how and why the Hubble expansion of the universe may appear locally. We discuss the observational aspects of this local cosmological effect in next section in more detail.

Solution of \eqref{v3x7} is given by quadratic function of time
\be\lab{oi6v}
x^\a=x^\a_0+k^\a\left(t+\frac{H}2t^2\right)\;,
\ee
where $x^\a_0$ is position of photon at time $t=0$, and $k^\a=(1,k^i)$ is a constant null vector with the unit vector $k^i$ pointing out in the direction of propagation of light. The reader may notice that the coordinate speed of light, $v^\a=k^\a(1+Ht)$, exceeds the fundamental value of $c=1$ for $t >0$. There is no violation of special relativity here because this effect is non-local - the speed is given with respect to the origin of the local coordinates. The local value of the speed of light measured at time $t$ at the current position of photon, is always equal to $c=1$. This is because the group of the conformal isometry includes the Poincare group as a sub-group which allows us to change the initial epoch and the initial position on the background manifold without changing the differential equation (\ref{v3x7}). 

Non-uniform propagation of light in the local frame may look counterintuitive as compared with our experience with special relativity. Nonetheless, this is how light propagates in the expanding universe. Equation \eqref{oi6v} is just a direct consequence of a standard light propagation formula in cosmology which reads in the global conformal coordinates, $y^\a=k^\a\eta$, where $k^\a=(1,k^i)$ is the null vector \cite{weinberg_1972}. Taking this law of propagation of light, substituting it to equations of the coordinate transformations \eqref{c4dd},\eqref{b6ft} and accounting for constant translation $x^\a_0$, we arrive at \eqref{oi6v} as expected. The non-uniform propagation of light in the local frame can be observed in the solar system, thus, making it possible to measure the Hubble expansion rate locally as contrasted to the cosmological observations of distant quasars.

\section{Cosmological effects in the local frame}\lab{rr1}

\subsection{Radar and laser ranging}\lab{d4d}
Precise dynamical modelling of orbital and rotational motion of astronomical bodies in the solar system (major and minor planets, asteroids, spacecraft, etc..) is inconceivable without radar and laser ranging. The ranging is an integral part of the experimental testing of general relativity and alternative theories of gravity in the solar system \cite{2006LRR.....9....3W,kopeikin_2011book,Williams_2012}. We are to check if the Hubble expansion can be measured in the ranging experiments.

Equation of light propagation in the local Gaussian coordinates $x^\a$ is given by \eqref{v3x7}. 
Let us consider radial propagation of light. The radial (always positive) spatial coordinate of photon is, $r=\sqrt{\d_{ij}x^i x^j}$. Let a light pulse be emitted at time $t_0$ at point $r_0$, reaches the target at radial coordinate $r>r_0$ at time $t$, and is immediately retransmitted to the point of observation being at radial distance $r_1<r$ to which it arrives at time $t_1$.  Propagation of the outgoing and incoming light rays are obtained from (\ref{oi6v}) where we demand that at the time of emission, $t_0$, the coordinate speed of light $\dot r(t_0)=1$ for both outgoing and incoming light rays. Equation of propagation for outgoing light ray is
\be\lab{tt8a}
r=r_0+(t-t_0)+\frac{H}2\left(t-t_0\ri)^2\;,
\ee
and propagation of the incoming light ray is described by
\be\lab{tt8b}
r_1=r\phantom{_0}-(t_1-t)-\frac{H}2 \left[(t_1-t_0)^2-(t-t_0)^2\right]\;.
\ee

Let us assume for simplicity that the radar ranging is conducted by the Hubble observer at the origin of the local coordinates so that both the points of emission and observation of the light signal are at the origin and have the radial coordinate, $r_0=r_1=0$. 
We define the {\it radar distance} by a standard equation \cite{LanLif,kopeikin_2011book}
\be\lab{tt9}
\ell\equiv\frac12\le(t_1-t_0\ri)\;,
\ee
which is a relativistic invariant due to the covariant nature of the proper time $t$ and the constancy of the fundamental speed $c=1$ in the geometrized system of units adopted in the present paper. After solving (\ref{tt8a}), (\ref{tt8b}) we obtain 
\ba\lab{tt9a}
t&=&\frac12\left(t_0+t_1\ri)+\frac12Hr^2\;,\\
\nonumber\\
\lab{tzi9}
\ell&=&r-Hr^2\;,
\ea
where the residual terms of the order of $O(H^2)$ have been neglected, $r$ is the radial distance of the point of reflection of a radar signal at time $t$. 

This calculation reveals that the difference between the coordinate distance $r$ and the invariant {\it radar distance} $\ell$ is of the order of $Hr^2$. Planetary ranging is done for the inner planets of the solar system so we can approximate $r\simeq 1$ astronomical unit (au) and $H\simeq 2.3\times 10^{-18}$ s$^{-1}$. Hence the difference $Hr^2\simeq 0.17$ mm which is a factor of $\sim 10^4$ smaller than the current ranging accuracy ($\sim 2$ m) to interplanetary spacecraft \cite{Folkner_2009IPNPR,Fienga_2009AA}. In case of lunar laser ranging to the Moon, the coordinate radius of the lunar orbit $r\simeq 384,000$ km, and the estimate of the residual term $Hr^2\simeq 1.1\times 10^{-6}$ mm which is one million times less than the current accuracy ($\sim$ 1 mm) of LLR \cite{murthyetal08}. We conclude that in radar/laser ranging experiments:
\begin{enumerate}  
\item[(1)] within the measuring uncertainty the coordinate radial distance $r=\ell$,
\item[(2)] the radial distance $r$ in the local frame of reference has an invariant geometric meaning in agreement with the definition of the proper distance accepted in cosmology \cite{weinberg_1972,mukh_book},
\item[(3)] the radar/laser ranging metrology is insensitive to the Hubble expansion in the local coordinates.  
\end{enumerate} 
Hence, the celestial ephemerides of the solar system bodies built on the basis of radar/laser ranging data are not crippled by the Hubble expansion. They represent a dynamical reference frame with a fixed value of the astronomical unit (au) which is not changing in time and can be treated as a rigid ruler for measuring distances between celestial bodies within the solar system in accordance with a recent resolution of IAU General Assembly (Beijing 2012) on the meaning and value of astronomical unit \cite{2012IAUJD...7E..40C}.

\subsection{Einstein's synchronization of clocks}\lab{bub3}
Let us now consider the Einstein procedure of the synchronization of two clocks based on exchange of light signals between the clocks. We want to synchronize the clock of the central Hubble observer with the clock of a static observer located at a point with the Gaussian radial coordinate $r$. We apply exactly the same procedure as in the case of radar ranging described above. By Einstein's definition, when the photon riches the reflection point with the radial coordinate $r$ at the instant of time $t$, the clock of the Hubble observer at the point, $r=0$, reads the time
\be\lab{tt10}
t^*=\frac12\le(t_0+t_1\ri)\;,
\ee
because the time rate of the (ideal) clock of the Hubble observer is uniform. The instant of time $t^*$ is defined as being simultaneous with the time reading, $t$, of a second clock located at the position with a radial coordinate, $r$, at the instant when the light signal is reflected. The time $t^*$ as a function of $t$, can be found immediately from (\ref{tt9a})
\be\lab{tt10a}
t^*=t-\frac12Hr^2\;.
\ee
This relation reveals that in order to synchronize two clocks separated by a radial distance $r$, we have to subtract the time difference $Hr^2$ from the reading $t$ of the clock of the static observer at the point with radial coordinate $r$ in order to make the time readings of the two clocks identical. Because the radial distance $r$ coincides with the invariant radar distance $\ell$, which is a measurable quantity, the Einstein synchronization of clocks in such experiment is operationally possible.

The two clocks will remain synchronized as time goes on, if and only if, the radial distance between the clocks does not change. For example,  a clock at a geocenter will remain synchronized with clocks on-board of a geostationary satellite moving around Earth on a circular orbit. On the other hand, an ultra-stable clock on board of spacecraft which moves with respect to the primary time standard on Earth may detect the de-synchronization effect due to the Hubble expansion of the universe if the radial distance between Earth and the spacecraft changes periodically. If the change in the radial distance amounts to $\d r$, the overall periodic time difference caused by the clock's de-synchronization amounts to $\d t=\d(t^*-t)=2H(r/c)^2(\d r/r)$. Expressing $r$ in astronomical units we can find a numerical estimate of the de-synchronization between the readings of the two clocks, 
\be\lab{tt10b}
\d t=1.7\times 10^{-12}\le(\frac{r}{1\;{\rm au}}\ri)^2\left(\frac{\d r}{r}\ri) \;\mbox{[s]}\;,
\ee
where we have used the approximate numerical value of the Hubble constant, $H=2.3\times 10^{-18}$ s$^{-1}$, the universal speed $c=3\times 10^{10}$ cm/s in cgs units, and the astronomical unit (au) $=1.5\times 10^{13}$ cm. This local cosmological effect may be detectable by NIST and/or other world-leading timekeepers.
\subsection{Doppler effect in the local frame}\lab{glb}

Next step is to consider the Doppler effect that is a change in frequency of propagating electromagnetic wave (light) emitted at one spacetime event and received at another one, as caused by various physical reasons - relative motion of observer and the source of light, gravity field, expansion of the Universe, etc. A monochromatic electromagnetic wave propagates on a light cone hypersurface of a constant phase $\varphi$, that is a function of spacetime coordinates, $\varphi=\varphi(x^\a)$. The wave (co)vector is $l_\a=\pd_\a\varphi$, and frequency of the wave measured by an observer moving with 4-velocity, $u^\a$, is \cite{waldorf,kopeikin_2011book}
\be\lab{d1}
\omega=-l_\a u^\a\;.
\ee
Frequency of electromagnetic wave can be calculated directly as soon as we know $l_\a$ and $u^\a=dx^\a/d\t$ where $\t$ is the proper time along the worldline of emitter (or receiver) of light. Indeed, 
\be\lab{n4d5}
\omega=-l_a u^\a=\frac{\pd\varphi}{\pd x^\a}\frac{dx^a}{d\t}=\frac{d\varphi}{d\t}\;,
\ee
which is just the rate of change of the phase of electromagnetic wave along the world line of emitter (or receiver).
 
Let us denote the point of emission of the wave $P_1$, the point of its observation $P_2$, and the emitted and observed wave frequencies as $\omega_1$ and $\omega_2$, respectively. Proper time of the emitter is denoted as $\t_1$, the proper time of receiver is $\t_2$, and the time measured by the central Hubble observer is the cosmic time $t$.
The ratio of received to emitted frequency 
\be\lab{d2}
\frac{\omega_2}{\omega_1}=\frac{\le(l_\a u^\a\ri)_{P_2}}{\le(l_\a u^\a\ri)_{P_1}}\;,
\ee
quantifies the Doppler effect. 

Because the phase of electromagnetic wave remains constant along the light rays we can use equation \eqref{n4d5} to reformulate \eqref{d2} in terms of the time derivatives \cite{kopeikin_2011book}
\be\lab{d2ss}
\frac{\omega_2}{\omega_1}=\frac{d\t_1}{dt_1}\frac{dt_1}{dt_2}\frac{dt_2}{d\t_2}\;,
\ee
where $d\t_1/dt_1$ is calculated along the worldline of emitter, $dt_2/d\t_2$ is calculated along the worldline of receiver, and $dt_1/dt_2$ is calculated along the light ray which is a solution of the light ray geodesic \eqref{v3x7} described by equation
\be\lab{b3c6y}
x^i(t)=x^i_0+k^i\left[(t-t_0)+\frac{H}2(t-t_0)^2\right]\;,
\ee
where $x^i_0$ is the point of emission of light at time $t_0$, and $x^i$ is the point of reception of light at time $t$, and we fix the coordinate speed of light $\dot x^i=1$ at time $t=t_0$.

We assume that the emitter and receiver moves with respect to LIC, and their proper times $\t_1$ and $\t_2$ are related to time $t_1$ and $t_2$ by the definition of the spacetime interval: $-d\t^2=ds^2=-(dx^0)^2+d{\bm x}^2$. Taking derivatives yield,
\ba\lab{aa1az}
\frac{d\t_1}{dt_1}&=&\sqrt{1-v^2_1+2H{\bm v}_1\cdot{\bm x}_1}\;,\\\lab{aa1ag}
\frac{d\t_2}{dt_2}&=&\sqrt{1-v^2_2+2H{\bm v}_2\cdot{\bm x}_2}\;,
\ea
where we have made use of diffeomorphism \eqref{c4} taken on the worldline of emitter and receiver respectively, and $v^i_1=dx^i_1/dt_1$, $v^i_2=dx^i_2/dt_2$ are velocities of the emitter and the receiver. 

Time derivative $dt_1/dt_2$ can be found from \eqref{b3c6y}. We write this equation for times $t_1$ and $t_2$, subtract one from another, and find the radial distance between points $x^i_1=x^i(t_1)$ and $x^i_2=x^i(t_2)$. It yields
\be\lab{u7b4}
|{\bm x}_2-{\bm x}_1|=t_2-t_1+\frac{H}2\left[\left(t_2-t_0\right)^2-\left(t_1-t_0\right)^2\right]\;.
\ee
Taking differential of this equation and separating terms being proportional to $dt_1$ and $dt_2$, we get
\be\lab{o7n4}
\frac{dt_1}{dt_2}=\frac{1-{\bm n}_{21}\cdot{\bm v}_2+H(t_2-t_0)}{1-{\bm n}_{21}\cdot{\bm v}_1+H(t_1-t_0)}
\ee
where the unit vector 
\be\lab{t5f4}
{\bm n}_{21}=\frac{{\bm x}_2-{\bm x}_1}{|{\bm x}_2-{\bm x}_1|}\;,
\ee
and it points out from the point of emission, $P_1$, to the point of reception, $P_2$, of the light signal.

We insert equations \eqref{aa1az}, \eqref{aa1ag}, \eqref{o7n4} to \eqref{d2ss}, and expand with respect to the Hubble constant. It yields the Doppler shift of frequency of electromagnetic wave in the expanding universe for the emitter and receiver being moving with respect to the LIC,
\be\lab{d5aqq}
\frac{\omega_2}{\omega_1}=\frac{1-{\bm n}_{21}\cdot{\bm v}_2}{1-{\bm n}_{21}\cdot{\bm v}_1}\sqrt{\frac{1-v_1^2}{1-v_2^2}}\Bigl[1+H(t_2-t_1)\Bigr]\;,
\ee
where we have dropped off all residual terms of the order of $Hv_1$ and $Hv_2$ as negligibly small. Notice that \eqref{d5aqq} does not depend on the choice of the initial epoch $t_0$. 

Equation \eqref{d5aqq} consists of two groups of term. The first group depend on velocities of emitter and receiver, and represents a special relativistic Doppler effect. The second group (in square brackets) depends on the Hubble constant $H$ and represents an additional shift of frequency caused by the cosmological expansion of space. Gravitational field of the solar system bodies should be also taken into account in realistic experiments. We have excluded the gravitational shift of frequency as it brings about much more terms to \eqref{d5aqq} and makes it too complicated. These terms are well-known and can be found, for example, in \cite{odprogram,kopeikin_2011book}.

For static emitter and receiver we have ${\bm v}_1={\bm v}_2=0$, and the Doppler shift equation \eqref{d5aqq} drastically simplifies
\be\lab{d5a}
\frac{\omega_2}{\omega_1}=1+H(t_2-t_1)\;.
\ee
It tells us that the cosmological Doppler shift measured by the local static observers is {\it blue} because $t_2>t_1$ and, consequently, $\o_2>\o_1$. It works opposite to the cosmological {\it red} shift for distant quasars \cite{weinberg_2008} but there is no contradiction over here. Cosmological red shift is measured with respect to the reference objects (quasars)
which have fixed values of the global coordinates, $y^i$, while the local Doppler shift (\ref{d5a}) is measured with respect to static observers having fixed Gaussian coordinates $x^i$. Thus,
the Doppler shift measurements in the global cosmological spacetime and in the local tangent spacetime refer to two different sets of reference observers moving one with respect to another with the velocity of the Hubble flow.
Therefore, it is natural to expect a different signature of the Doppler effect -- {\it red} shift for light coming from distant quasars and {\it blue} shift for light emitted by the astronomical objects, for example spacecraft, within the solar system. Our theory provides an exact answer for the signature and magnitude of the cosmological {\it blue} shift effect measured in the local inertial frame. 

The Doppler effect in the tangent spacetime of FLRW universe has been considered by a number of other authors, most notably by Carerra and Giulini \cite{2006CQGra..23.7483C,2010RvMP...82..169C}. They claimed that the cosmological expansion does not produce any Doppler effect in the local radio-wave frequency measurements. Their conclusion is invalid as they implicitly identified the local Minkowskian time coordinates $x^0$ with the proper time $t$ of the Hubble observer on a worldline of any freely-moving particle including photons. However, this identification is not applied to photons (or any other moving particle) but solely to the static clocks of the Hubble observer. This is the reason for the overlook admitted in \cite{2006CQGra..23.7483C,2010RvMP...82..169C}.

\subsection{Measuring the Hubble constant with spacecraft Doppler-tracking}\lab{mdt5nc}

Results of previous section suggest that precise and long-term Doppler tracking of space probes in the solar system may offer a new, fascinating opportunity to measure the local value of the Hubble constant $H$ in the solar system. It is highly plausible that the ``Pioneer anomaly'' detected by John Anderson \cite{2002PhRvD..65h2004A} with the JPL deep-space Doppler tracking technique in the hyperbolic orbital motion of Pioneer spacecraft has a natural explanation given in terms of the Hubble expansion which changes the frequency of radio waves in spacecraft radio communication link in an amazing agreement (both in sign and in magnitude) with our equation (\ref{d5a}). 

We have analyzed the cosmological origin of the ``Pioneer anomaly'' effect in another paper \cite{Kopeikin_2012eph} making use of the local equations of motion for charged and neutral test particles as well as for photons in FLRW universe. We have proved that in the local frame of reference the equations of motion for interacting massive neutral and/or charged particles do not include the linear terms of the first order in the Hubble constant -- only tidal terms of the order of $H^2$ remain. On the other hand, equations of motion of photons parameterized with the TCB time $t$ do contain such linear terms of the order of $H$ which have dimension of acceleration. 

The present paper confirms results of the paper \cite{Kopeikin_2012eph} from the point of view of a set of local observers doing measurements in tangent space of the FLRW manifold. Transformation to the local coordinates $x^\a=(x^0,x^i)$ allows us to transform FLRW metric to the Minkowski metric $ds^2=-(dx^0)^2+\d_{ij}dx^idx^j$ but the coordinate time $x^0$ can be identified with the proper time $t$ of the central Hubble observer only for static observers while for moving particles $x^0=x^0(\t)$ is a non-linear function of time $t$ which is given for photons by \eqref{yu6v}.

Equation (\ref{d5aqq}) explains the ``Pioneer anomaly'' effect as a consequence of the expansion of space bringing about the {\it blue} frequency shift of radio waves on their round trip from Earth to spacecraft and back. Indeed, let us denote $\o_0$ -- the reference frequency emitted from Earth to spacecraft,  $\o_1$ -- the frequency received at spacecraft and transmitted back to Earth, and  $\o_2$ -- the frequency received on Earth. Then, according to (\ref{d5aqq}), the shift between $\o_0$ and $\o_2$ is
\ba\lab{fgl9}
\frac{\o_2}{\o_0}=\frac{\o_2}{\o_1}\frac{\o_1}{\o_0}&=&\frac{(1-{\bm n}_{21}\cdot{\bm v}_2)(1-{\bm n}_{10}\cdot{\bm v}_1)}{(1-{\bm n}_{21}\cdot{\bm v}_1)(1-{\bm n}_{10}\cdot{\bm v}_0)}\\\nonumber
&\times&\sqrt{\frac{1-v_0^2}{1-v_2^2}}\Bigl[1+H(t_2-t_0)\Bigr]\;,
\ea
where $t_0$ is the time of emission on Earth, $t_1$ is the time of re-transmission of the signal at spacecraft, and $t_2$ is the time of reception of the signal back on Earth.

Let us simplify further consideration by assuming that the measurement is done by the central Hubble observer located at the origin of LIC. Then, ${\bm v}_0={\bm v}_2=0$, and the unit vector ${\bm n}\equiv{\bm n}_{10}=-{\bm n}_{21}$ points out in the positive radial direction toward spacecraft moving with velocity ${\bm v}\equiv{\bm v}_1$. After noticing that $t_2-t_0\simeq 2(t_1-t_0)$ and neglecting quadratic with respect to velocity terms, formula \eqref{fgl9} takes on the following form
\be\lab{g7n0}
\frac{\o_2}{\o_0}=1-2[\frac{v}c-H(t_1-t_0)]\;,
\ee
where $v={\bm n}\cdot{\bm v}$ is the radial velocity of spacecraft, and we prefer to keep over here the speed of light $c$ explicitly. The Doppler shift is defined as $z\equiv (1/2)[(\o_2/\o_0)-1]$. We get from  \eqref{g7n0} 
\be\lab{sxw2}
z=-\frac{v}c+H(t_1-t_0)\;,
\ee
that shows that the cosmological shift of frequency appears as a tiny {\it blue} shift on top of much larger {\it red} shift of frequency caused by the outward motion of the spacecraft. It was observed by J. Anderson \cite{2002PhRvD..65h2004A} and confirmed in a number of papers \cite{2007AdSpR..39..291T,2011AnP...523..439R}.

The time rate of change of the Doppler shift is $\dot z\equiv dz/dt_1$ which yields
\be\lab{srb6}
\dot z=-\frac1c\le(a-Hc\ri)\;,
\ee
where $a=dv/dt_1$ is the magnitude of radial acceleration of the Pioneer spacecraft due to the attraction of the solar gravity field. The Hubble frequency-shift term, $Hc$, is subtracted from the spacecraft acceleration and can be interpreted as a constant, directed-inward acceleration, $a_P=Hc$, in the motion of spacecraft. In fact, the true cause of the ``anomalous'' acceleration is associated with the motion of photons but not the spacecraft. This is the reason why the vigorous attempts to find out the explanation for the ``anomalous gravity force'' exerted on Pioneer spacecraft were unsuccessful. The observed value of $a^{\rm obs}_P=8.5\times 10^{-10}$ m$\cdot$s$^{-2}$ \cite{2002PhRvD..65h2004A} is in a good agreement (both in sign and in magnitude) with the theoretical value of $a^{\rm theory}_P=Hc\simeq 7\times 10^{-10}$ m$\cdot$s$^{-2}$. Therefore, we believe that our result \eqref{srb6} provides a strong evidence in favour of general-relativistic explanation of the ``Pioneer anomaly'' as opposed to numerous attempts to explain it by thermal recoil force.  

The thermal recoil definitely makes contribution to the acceleration of Pioneer spacecraft because the observed value of $a^{\rm obs}_P$ exceeds theoretical value $a^{\rm theory}_P$ by 20\% \cite{2002PhRvD..65h2004A,2007AdSpR..39..291T}. Recent studies \cite{2011AnP...523..439R,Turyshev_2012,2013arXiv1311.4978M} indicate that the numerical value of the Pioneer anomalous acceleration may be slightly decreasing over time which may be associated with the radioactive decay of the power generators of Pioneer spacecraft. The question about how much the thermal recoil force contributes to the overall effect remains open. The papers  \cite{2011AnP...523..439R,Turyshev_2012,2013arXiv1311.4978M} state the Pioneer effect is 100\% thermal but they have not taken into account the geometric effect of the expanding space on the propagation of light in the local frames in cosmology which suggests that numerical value of the Pioneer effect cannot be smaller than $a^{\rm theory}_P=7\times 10^{-10}$ m$\cdot$s$^{-2}$. The thermal emission always adds to the general-relativistic prediction, $a^{\rm theory}_P$. Observations indeed show $a^{\rm obs}_P$ larger than $a^{\rm theory}_P$ by 20\%. The theory of the present paper explains 80\% of the overall effect by the effect of the expanding geometry leaving for the thermal recoil contribution no more than 20\%.

\section{Discussion}\lab{opwd}
\begin{enumerate}
\item We have build the LIC by applying the special conformal transformation (\ref{c4}). Comparison with other approaches \cite{hongya:1920,hongya:1924,2005ESASP.576..305K,2007CQGra..24.5031M,2010RvMP...82..169C} to build the LIC in cosmology reveals that all of them bring about the same coordinate transformation \eqref{c4rr}, \eqref{c4sf} in the linearized Hubble approximation. Therefore, there is no difference between various approaches to build the local inertial coordinates in cosmology so far as the quadratic terms in the expansion with respect to the Hubble parameter are not considered. Our approach to build LIC helps to realize that the transformation to the local coordinates on the expanding cosmological manifold is, in fact, an infinitesimal special conformal transformation which establishes 1-to-1 local mapping between the local and conformal coordinates.    

\item  Introducing a local physical distance $x^i=R(t)y^i$, recast (\ref{1}) to the following form
\be\lab{mzd4}
ds^2=-\le(1-H^2{\bm x}^2\ri)dt^2-2Hx^idx^idt+\d_{ij}dx^idx^j\;,
\ee
which can be written down as
\be\lab{mb3df}
ds^2=-dt^2+\d_{ij}\left(dx^i-X^idt\ri)\left(dx^j-X^jdt\ri)\;,
\ee
where vector field $X^i\equiv Hx^i$. Metric (\ref{mb3df}) is exactly the warp-drive metric that was suggested by Alcubierre \cite{warp_drive} to circumvent the light-speed limit in general relativity. All mathematical properties of the warp-drive metric that have been analysed, for example in \cite{warp_drive2002}, are valid in the local coordinates $(t,x^i)$ where $t$ is the proper time of the local static observers ($x^i={\rm const.}$) coinciding with the cosmic time. The metric (\ref{mb3df}) is non-inertial but it can be converted to the flat Minkowski metric in a neighbourhood of the coordinate origin with the help of an additional transformation of the proper time $t$ to a local time coordinate $x^0$ as shown in (\ref{c4}). The local time coordinate $x^0$ coincides with the proper time $t$ of the static observers but deviates quadratically from $t$ on the light cone as demonstrated in \eqref{yu6v}. 

\item The analysis of EEP given in the present paper, was focused on the solar system experiments as contrasted with pure cosmological tests. There are other possible tests which can be potentially conducted for testing the formalism worked out in the present paper, for example, with binary pulsars \cite{krawex_2009}. Timing measurements establish a very precise local frame for the binary pulsar system which is not affected by the Hubble expansion as explained in \cite{Kopeikin_2012eph}. On the other hand, we expect that the cosmological expansion influence the time of propagation of radio pulses from the pulsar to observer on Earth, and this effect should be seen in the secular change of the orbital period $P_b$ of binary pulsars of the order of $\dot P_b/P_b=H\simeq 2.3\times 10^{-18}$. This effect is superimposed on the effect of the orbital decay due to the emission of gravitational waves by the binary system and introduces a bias to the observed value of $\dot P_b$ in addition to the Shklovskii effect \cite{Damour_1991ApJ}. However, the orbital decay of binary pulsars with wide orbits is negligible small, hence, we may expect to observe the Hubble expansion effect in the secular change of the orbital period. 

\item It is worth mentioning that Cassini spacecraft was also equipped with a coherent Doppler tracking system and it might be tempting to use the Cassini telemetry to measure the universal ``anomalous Cassini acceleration'' $a^{\rm theory}_C\simeq 7 \times 10^{-10}$ m$\cdot$s$^{-2}$. Unfortunately, there are large thermal and outgassing effects on Cassini that would make it difficult or impossible to say anything about the ``Cassini anomaly'' from Cassini data, during its cruise phase between Earth and Saturn \cite{2010PEPI..178..176A}. Due to the presence of the Cassini-on-board-generated systematics, the recent study \cite{2012CQGra..29w5027H} of radio science simulations in general relativity and in alternative theories of gravity is consistent with a non-detection of the ``Cassini anomalous acceleration'' effect.
\end{enumerate}

\begin{acknowledgement}
I am grateful to S. Alighieri, Yu. Baryshev, M. Khlopov, S. Levshakov, B. Mashhoon, M. Soffel and O. Titov for valuable discussions and critical comments. The present work has been supported by the Faculty Fellowship 2014 in the College of Arts and Science of the University of Missouri and the grant 14-27-00068 of the Russian Scientific Foundation. 
\end{acknowledgement}
\newpage
\bibliographystyle{unsrt}
\bibliography{kopeikin_bib}
\addcontentsline{toc}{section}{\hspace{0.55cm}References}
\end{document}